\documentclass[letterpaper, 10 pt, conference]{ieeeconf}
\pdfoutput=1
\usepackage{amsmath}
\usepackage{amssymb}

\usepackage{subcaption}

\IEEEoverridecommandlockouts                             
\usepackage{graphicx}

\title{Reinforcement Learning Approach to Vibration Compensation for Dynamic Feed Drive Systems}

\author{Ralf Gulde, Marc Tuscher, Akos Csiszar,  Oliver Riedel and Alexander Verl
\thanks{R. Gulde, M. Tuscher, A. Csiszar, O. Riedel and A. Verl
are   with   the   Institute   of   Control   Engineering   of   Machine   Tools   and   
Manufacturing  Units,  University  of  Stuttgart,  70174  Stuttgart,  Germany  
(phone:      0049-711-685-82505;      fax:      0049-711-685-72505;      e-mail:      
ralf.gulde@isw.uni-stuttgart.de). The Research is supported by the Graduate School of Excellence advanced Manufacturing Engineering and by the German Research Foundation (DFG).}}

\begin{document}

\maketitle
\thispagestyle{empty}
\pagestyle{empty}

\begin{abstract}
Vibration compensation is important for many domains. For the machine tool industry it translates to higher machining precision and longer component lifetime. Current methods for vibration damping have their shortcomings (e.g. need for accurate dynamic models). In this paper we present a reinforcement learning based approach to vibration compensation applied to a machine tool axis. The work describes the problem formulation, the solution, the implementation and experiments using industrial machine tool hardware and control system.

\end{abstract}

\section{Introduction}
The compensation of mechanical and structural vibration has significant applications in manufacturing, infrastructure engineering and other domains. 
In automotive or aerospace applications, vibration reduces component lifetime and the associated acoustic noise can produce discomfort. In machine tools residual vibrations degrade the position accuracy and produce material fatigue. 

The compensation of such mechanical vibrations is a large and important field of research. Various methods have been applied to provide solutions for this challenging problem.

In this paper we briefly review the state of the art in the field of vibration compensation of dynamic feed drive systems (the main sources of motion in machine tools) and describe drawbacks in the solutions provided in the state of the art. 
We propose a novel approach based on deep reinforcement learning to compensate vibrations in dynamic drive systems with a priori unknown system parameters. 
The proposed method is experimentally validated using a linear direct drive and control hard- and software customary in the machine tool industry.
\section{State of the Art}

The research in the field of vibration compensation can roughly be broken into three categories: hardware design, command shaping, and feedback control. The above-mentioned research areas are illustrated in the followings.

\subsection{Hardware Design} 
In hardware design approaches, vibration compensation is achieved by using additional mechanical systems. The damping of the system is increased by mass dampers and vibration absorbers. The advantage of these systems resides in their simple construction and cost-effective implementation. The major drawback in hardware design approaches is the low flexibility \cite{weck2006werkzeugmaschinen} since hardware (e.g. passive dampers \cite{soong1997passive}) is used to overcome an application specific problem.

\subsection{Command Shaping}
Command shaping methods are altering the reference motion trajectory in order to filter out specific frequencies. Thus the suppression of vibrations is done in a preemptive way.
The major downside of command shaping methods is that a dynamic model of the system or at least its natural frequencies and damping behavior has to be known beforehand with sufficient accuracy. The dynamic model has to be re-evaluated when the system parameters vary.
One of the earliest publications on command shaping is \cite{smith1957posicast}. Smith et al. proposed a method, known as posicast control, that processes a baseline command and delay a part of the command before transferring it to the system. The delayed portion of the command canceles out the vibration induced by the undelayed part of the motion command. 
A key advancement in command shaping was the concept of robustness – commands can be designed to work well, even when large modeling errors exist. Singer and Seering presented an input-shaping method \cite{singer1990experimental} that increases the robustness of the input-shaping process. They used an additional constraint to enforce the derivative of the residual vibration, with respect to the frequency, to equal zero:

\begin{equation}
\frac{\partial}{\partial(\omega)}V(\omega, \xi) = 0.
\end{equation}

Where $\omega$ is the natural frequency, and $\xi$ is the damping ratio. When $V(\omega, \xi) = 0$ is satisfied, the result is a Zero Vibration and Derivative (ZVD) shaper containing three impulses.

\subsection{Feedback Control}
\begin{figure}[ht]
\centering
\includegraphics[width=0.70\columnwidth]{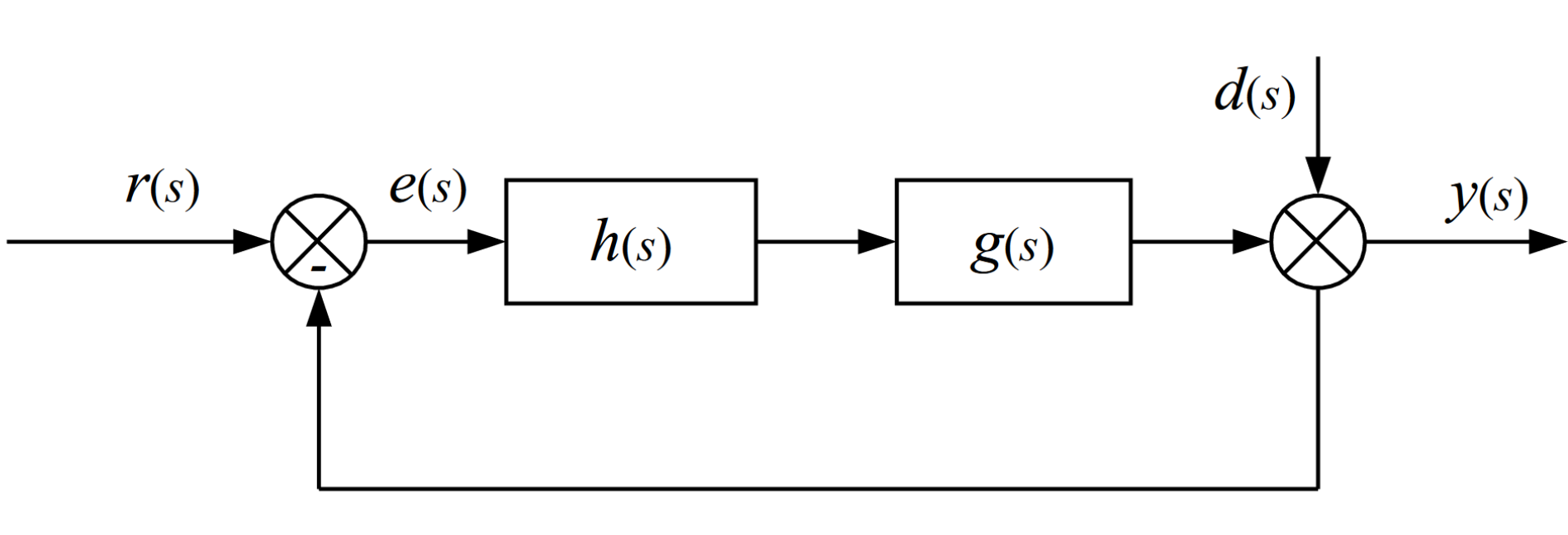}
\caption{Feedback control \cite{smith1958feedback}.}
\label{fig: feedback control}
\end{figure}

Vibration compensation using feedback control, also known as active vibration control, incorporates sensors to measure the mechanical disturbance, a controller to compute an appropriate counter-vibration and control an actuator accordingly. Destructive  interference  from  additional movements  generated  by  the  controller  reduces  or  neutralizes the effects of the disturbance on the structure \cite{smith1958feedback}. The scheme of feedback control is depicted in Figure (\ref{fig: feedback control}) . The feedback signal $e=r-y$ is computed from the comparison of the  output $y$ of the system and the input $r$. The error signal is passed into a compensator $h(s)$ and applied to the system $g(s)$. The controller is designed with the aim of determining an appropriate transfer function of the compensator $h(s)$, to induce the sought-after performance while maintaining the system stability. 

The objective of active damping is to reduce resonant peaks of the closed-loop control circuit

\begin{equation}
f(s)= \frac{y(s)}{r(s)}=\frac{g(s)h(s)}{1+g(s)h(s)}.
\end{equation}

Active damping can be achieved without modelling the system dynamics, but is only effective near resonance peaks. Moreover the stability can only be guaranteed when the sensors and actors are collocated. 
Model based methods attenuate all disturbance within the control bandwidth, but require an accurate model of the system. Such methods generally have limited bandwidths and cope with control and observation spillover \cite{meirovitch1987some}. 

Coordinate Coupling Control (CCC) is an energy-based method to eliminate the transient vibration of an oscillatory system \cite{golnaraghi1991regulation}. The technique was later extended to compensate steady-state vibrations 
\cite{ashour2002adaptive}. 

Robust control approaches focus the trade off between performance and stability in the presence of system model uncertainties. The $H_\infty$-controller is designed to address the uncertainties systematically. $H_\infty$-methods formulate the control problem as a mathematical optimization problem and solve it. The resulting $H_\infty$-controller is then optimal with respect to the prescribed cost function. Application of the $H_\infty$ method in the vibration control of flexible structures can be found in \cite{bayard1998identification, chang2002design}.

Optimal control theory applied in vibration control aims to reduce the vibration of the mechanical system to the greatest possible extent. The method seeks to compute the feedback gain by minimizing a cost function or a performance index, which is correlated to the required measure of the system response. Popular approaches are \cite{zhang2002optimal,  cai2002optimal}. 

There are attempts in the state of the art for reducing vibrations using machine learning. However, these in contrast to our proposed approach are not using reinforcement learning, like the neurofuzzy approach in \cite{Daniali2009Neurofzzy} or use reinforcement learning in simulation for active automotive suspension in simulation \cite{Bucak2012activesusp} and not for machine tools on industrial hardware.

The main shortcomings of the state of the art methods are the following problem statements:
\begin{itemize}
\item Complex modelling of the underlying system dynamics.
\item Learn from past performance to improve future actions.
\item Automatic adaption of the vibration compensation behavior
to changes in the structure/system.
\end{itemize}

\section{Preliminaries}

\subsection{Reinforcement Learning and Policy Optimization}
Further we define the Reinforcement Learning (RL) problem and introduce the notation that we use throughout the paper. 
In this paper a finite-horizon, discounted Markov Decision Process (MDP) is regarded. 
At each timestep $t$, the RL-agent observes the current state $s_t \in S$, performs an
action $a_t \in A$, and then receives a reward $r_{t+1} \in \mathbb{R}$. 
After that the resulting state $s_{t+1}$ will be observed, determined by the unknown dynamics of the environment $p(s_{t+1}| a_t, s_t)$. An episode has a pre-defined length $T$ time steps.
The goal of the agent is to find a parameter $\theta$ of a policy ${\pi_{\theta} (a|s)}$ that maximizes the expected cumulated reward $J$ over a trajectory
\begin{equation} \label{bell}
J(\pi_{\theta}) = \mathbb{E}_{\tau\sim{\pi_{\theta}}} \bigg[ \sum_{t=0}^T \pi(a_t | s_t) \sum_{k=t}^T \gamma^{k-t} r_{k+1} \bigg],
\end{equation}
where $\gamma \in [0, 1]$ is the discount factor. 

RL methods solve a MDP by interacting with the system and accumulating the obtained reward.
We consider several model-free policy gradient algorithms with open source implementations which appear frequently in the literature, e.g. Soft Actor-Critic approaches \cite{haarnoja2018soft}, Deep Deterministic Policy Gradient (DDPG) \cite{silver2014deterministic}, and Proximal Policy Optimization (PPO)~\cite{schulman2017proximal}.
The major advantage for the use of the PPO algorithm is that it allows to incorporate a Long Short-Term Memory (LSTM) \cite{hochreiter1997long} effortlessly~\cite{mnih2016asynchronous}. 
A LSTM is a specific recurrent neural network (RNN) architecture that was designed to model temporal sequences and their long-range dependencies more accurately than conventional RNNs~\cite{sak2014long}. 
The use of a LSTM significantly increases the model quality of the system dynamics (e.g. determining the actual vibrations from subsequent deflection observations). 
Therefore we use the PPO algorithm~\cite{schulman2017proximal} for the training of the agent. 

Generally, PPO maximizes (\ref{bell}) using a robust version of the policy gradient theorem
\begin{equation} \label{pg}
\nabla_{\theta} J(\pi_{\theta}) = \mathbb{E}_{\tau\sim{\pi_{\theta}}} \bigg[ \sum_{t=0}^T \nabla_{\theta} \log \pi(a_t | s_t) \sum_{k=t}^T \gamma^{k-t} r_{k+1} \bigg]
\end{equation}
and performing gradient ascent steps
\begin{equation}
    \theta_{k+1} = \theta_{k} + \alpha \nabla_{\theta}J(\pi_{\theta}).
\end{equation}

\begin{figure}[ht]
\centering
\includegraphics[width=0.99\columnwidth]{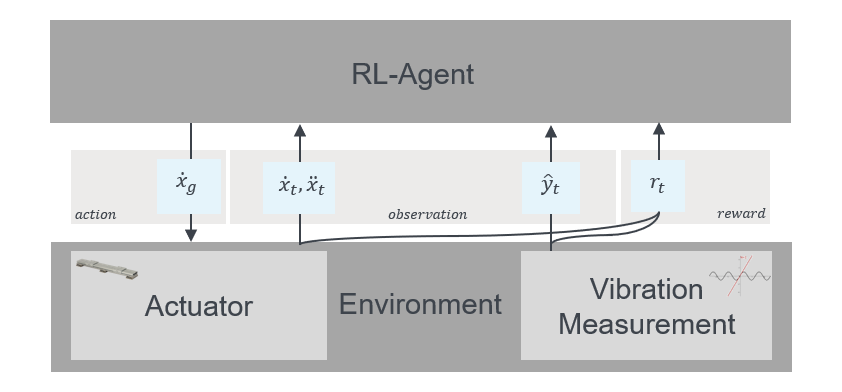}
\caption{System Architecture illustrated as MDP.}
\label{fig: arch}
\end{figure}

\begin{figure*}[t]
\centering
\begin{subfigure}[b]{\textwidth}
\centering
\includegraphics[width=\textwidth]{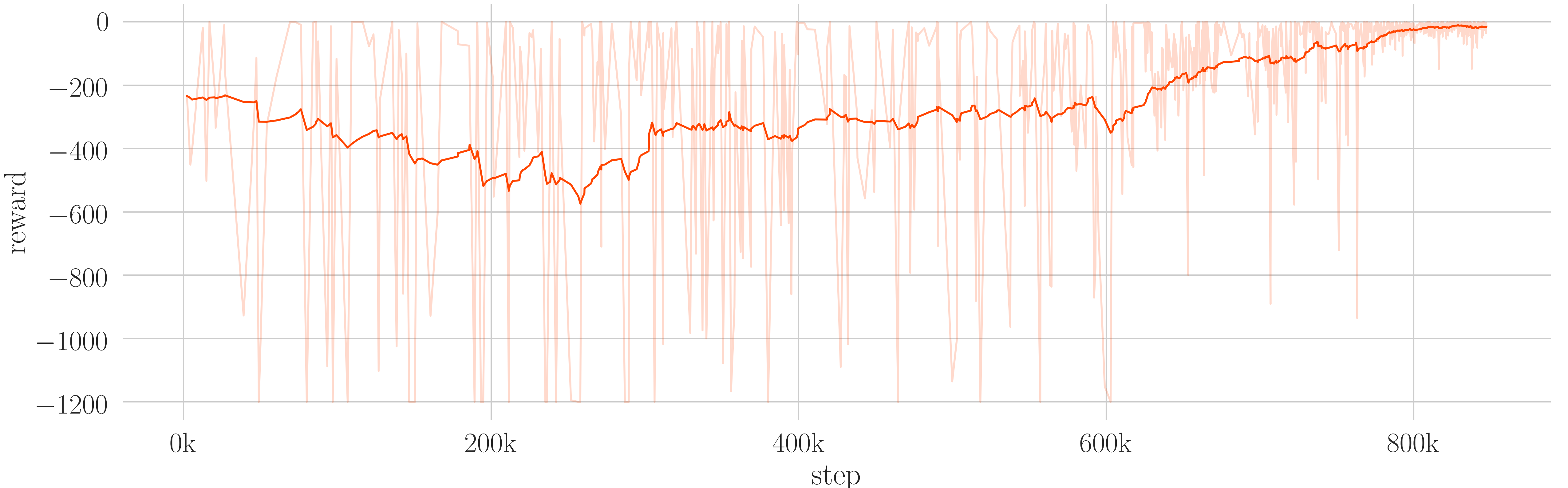}
\caption[]{\small}
\label{fig:a}
\end{subfigure}
\begin{subfigure}[b]{0.3\textwidth}
\centering
\includegraphics[width=\textwidth]{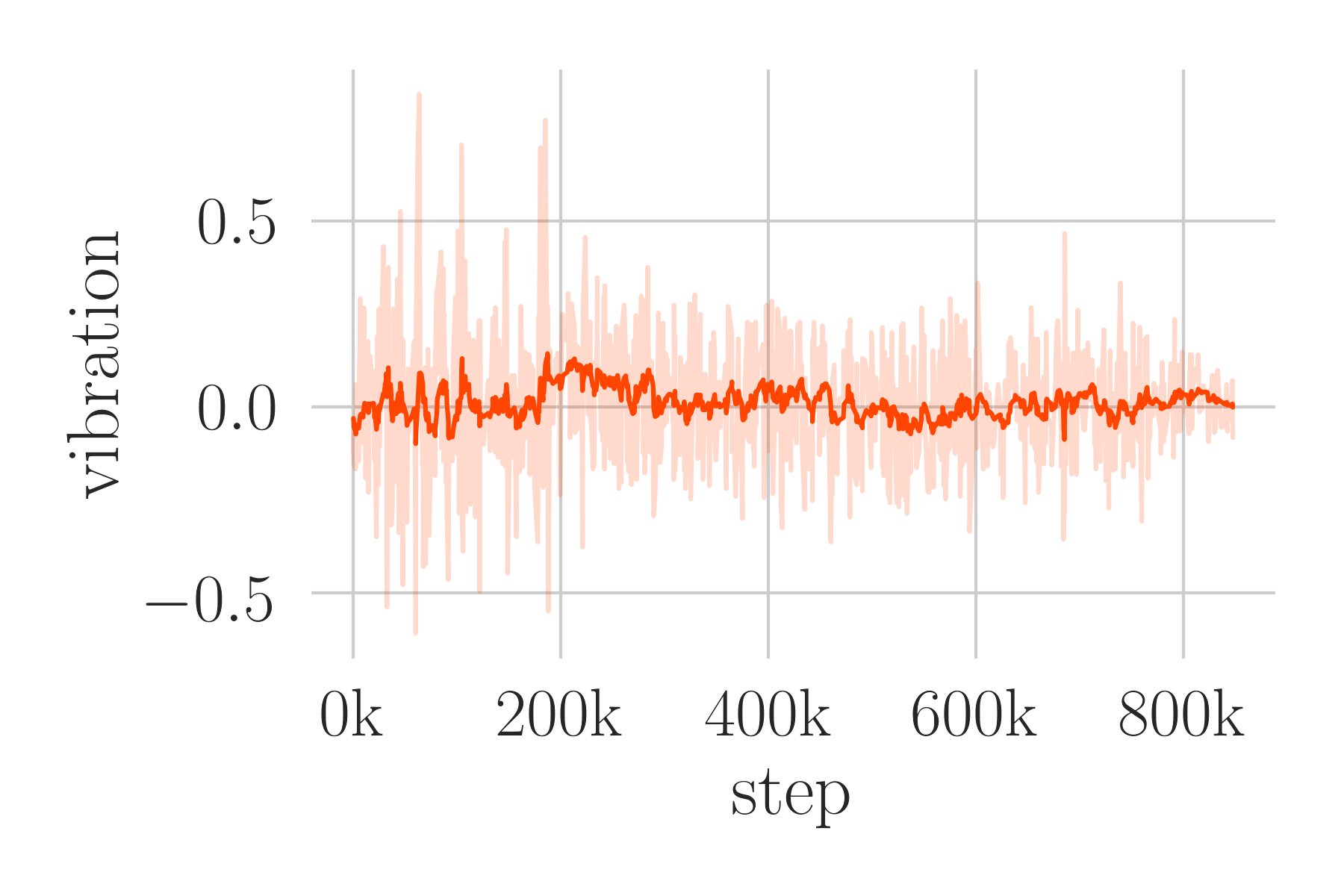}
\caption[]{\small}
\label{fig:b}
\end{subfigure}
\hfill
\begin{subfigure}[b]{0.3\textwidth}
\centering
\includegraphics[width=\textwidth]{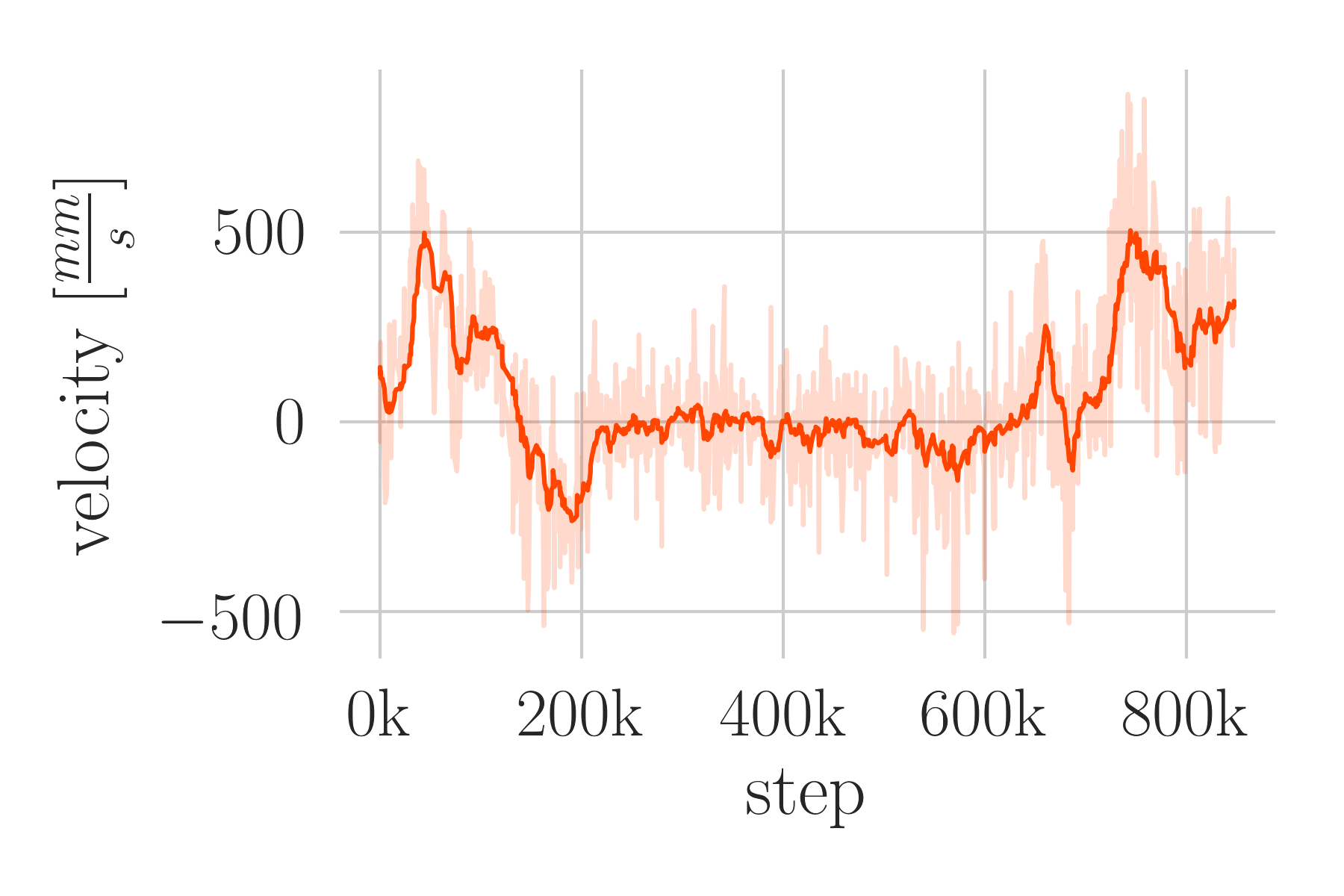}
\caption[]{\small}
\label{fig:c}
\end{subfigure}
\hfill
\begin{subfigure}[b]{.3\textwidth}
\centering
\includegraphics[width=\textwidth]{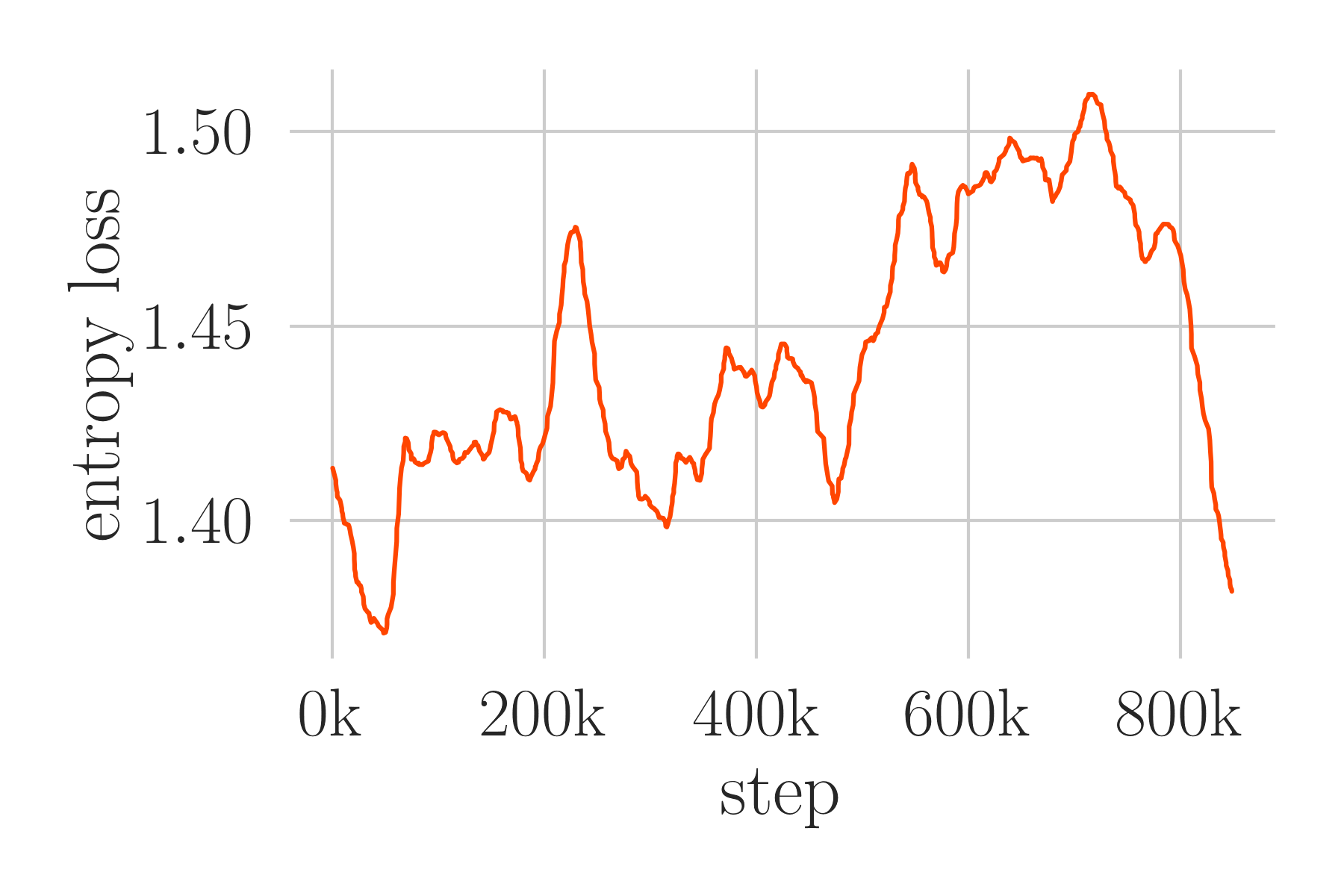}
\caption[]{\small}
\label{fig:d}
\end{subfigure}
\caption{Results of a training process plotted over all time steps(a): Illustrates the development of the episode-reward.  (b) shows the actual deflection of the vibration. (c): Depicts the velocity commands determined by the agent. (d): Represents the entropy loss.}
\vspace{-15pt}
\label{fig:experiments}
\end{figure*}
\section{Vibration Compensation using RL}
\subsection{Problem Formulation} \label{prob_form}
We consider vibration compensated movement commands that can be described as moving to a target position goal and compensating vibrations along the way.
Let $x_{t}$ and $y_{t}=\hat{y}\sin{(\omega t)}$ denote the actual position and the actual vibration, respectively, at time $t$ and let $\dot{x}_t$ denote the control command (velocity) applied to the system at that time. 
A movement command $s_{g} = ({x_{g},y_{g}})$ is described by a target position $x_g$ and a desired vibration $y_{g}$. To compensate occurring vibrations $y_{g}$ is set to 0. %$(0,0)^\mathsf{T}$. 
% By choosing $y_g\neq(0,0)^\mathsf{T}$, the desired oscillation is excited during execution of the motion command.
Given the movement command $s_{g}$, an initial system state $s_{0}=({x_{0},y_{0}})$, and a time horizon $T$ our vibration compensation problem is formulated as
\begin{align} 
\label{eq_loss}
\min_{\dot{x}_0,...,\dot{x}_T}  l(s_T, s_g)&, \\
 s_{t+1} = f(s_t, \dot{x}_t)&,~t \in 0,...,T.
\end{align}
Where $f$ describes the unknown dynamics of the system and $l$ the loss function defined by the summed squared distance
\begin{equation}
l(s_T, s_g) = \sum_{t}{{||s_t - s_g||}^2}.
\end{equation}

Generally, policy optimization approaches seek for the parameters $\theta$ of a reactive, fully parametrized policy ${\pi_{\theta} (\dot{x}_t|s_t)}$ such that selecting actions according to $\pi$ minimizes the loss in (\ref{eq_loss})~\cite{schulman2017proximal}.
\subsection{MDP Architecture}
The flow of information of our method is shown in figure~\ref{fig: arch}.
With respect to the introduced notation for the MDP we define observations, actions, and rewards as follows:

\textbf{Observation}. A state $s_t$ is described by the actual position $x_t$ and velocity $\dot{x}_t$ of the feed drive translator and the current vibration. Note that an observation of the current vibration solely incorporates the subsequent measurements of the deflection, not the frequency. The agent has to determine the actual frequency based on five preceeding measurements of the deflection $\{\hat{y}_i\}_{i=t-4,...,t}$.

\textbf{Action}. The agents action $a_t$ is defined by a continuous velocity command $\dot{x}_g \in [-400  \frac{mm}{s}, 400 \frac{mm}{s}]$.

\textbf{Reward}. The reward signal is described as follows
\begin{equation} \label{reward}
r_{t+1} =\left\{\begin{array}{ll}  ~~0, & (|\frac{x_t-x_g}{x_g}| + |\frac{\hat{y}_t-\hat{y}_g}{\hat{y}_g}|) < 0.01 \\
 -1, & otherwise\end{array}\right. ,
\end{equation}
Given the reward function in formula (\ref{reward}) a negative return is received while the target position $x_{g}$  and/or the desired vibration is not reached. Otherwise the agent receives $r_{t+1} =~0$.

For the RL agent a vibration-compensated motion is an opposing goal: Dynamically moving a machine axis induces vibrations; compensating vibrations affects the desired motion. 
This insight is used in the modelling of the reward function. We want the agent to fulfill both mutually influencing goals ($x_g$ and $y_g$). Therefore we design the reward function based on a sparse reward setting, treating positioning accuracy and vibration suppression equally. 
To learn from sparse rewards, effective exploration is crucial to find a set of successful trajectories. To guarantee sufficient amount of exploration we use the entropy coefficient as a regularizer. In a policy optimization setting, a policy has maximum entropy when all policies are equally likely and minimum when the one action probability of the policy is dominant. The entropy coefficient is multiplied by the maximum possible entropy and added to the loss and therefore prevents premature convergence of one action probability dominating the policy and preventing exploration \cite{schulman2017proximal}. Further, to ensure a high generalization performance of the agent, the target position is randomized during the training process.

\section{Experiments}

The experiments were done using a linear direct drive depicted in figure 4 coupled to a TwinCAT control unit. The RL Agent is deployed on a Ubuntu Xenial computer with a ADS (Automation Device Specification) interconnection to the control unit. Further, we use the Stable Baselines \cite{stableBaselines} implementations of RL algorithms. For hyperparameter tuning we apply a bayesian optimization approach provided by the framework optuna \cite{optuna}. To measure the mechanical vibrations we utilize a vision system using OpenCV.

\begin{figure}[ht]
\centering
\includegraphics[width=0.65\columnwidth]{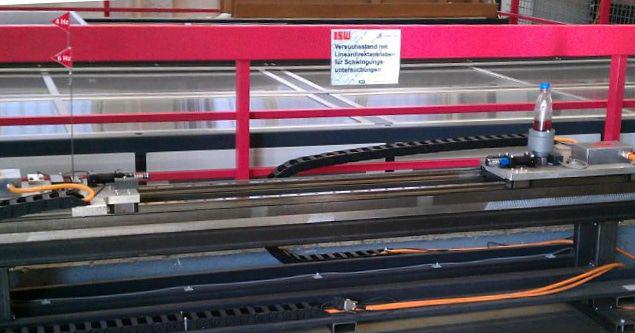}
\caption{Linear feed drive system.}
\label{fig: axis}
\end{figure}
Our experiment evaluates the cost function proposed in Section \ref{prob_form}. For this experiment, we want the linear feed drive to reach various, random sampled target positions and suppress vibrations along the way. Consequently we define the goal state as $s_{g}=(p_g\sim \mathcal{U}, 0)$ and the number of time steps $T$. Figure (\ref{fig:experiments}) illustrates the results. The agent solves the vibration compensation problem after 850.000 time steps, equalling 12 hours training on the real machine tool axis (cf. Figure (\ref{fig: axis})). Figure (\ref{fig:experiments}, a) shows the episode reword converging asymptotically towards zero after 850.000 time steps. Consequently the occurring vibrations (Figure \ref{fig:experiments}, b) also converge towards zero. Figure (\ref{fig:experiments}, d) illustrates the entropy loss that regularizes when the learning rate decays and attenuates when agents rewards converges.

\section{Conclusion}

In this work a reinforcement learning based approach to compensate mechanical vibrations applied to an industrial machine tool axis is presented. We propose a problem formulation describing the vibration compensation based on a vibration cost optimization problem. We evaluate different state of the art Reinforcement Learning algorithms to solve the vibration compensation problem. We train the agent directly on a real machine tool axis, without the use of a simulation environment. To validate our method we perform experiments on a real machine tool axis. The experiments show that the proposed approach is capable of generating vibration compensated movements using a feed drive system with a priori unknowns system dynamics. 

Further research could be conducted on the following topics: Deploy the agent using a discrete action space (move left; move right); investigate the generalization across varying machine tool hardware, utilize better vibration measurement system (more accurate and frequent observations).

\bibliography{root}{}

\end{document}